\documentclass[twocolumn] {jpsj2}

\def\runtitle{Observation of Superconductivity in Heavy-Fermion Compounds of Ce$_{2}$CoIn$_{8}$}
\def\runauthor{Genfu \textsc{Chen}\ {\it et al.}}

\title{%
Observation of Superconductivity\\in Heavy-Fermion Compounds of Ce$_{2}$CoIn$_{8}$
}

\author{%
Genfu \textsc{Chen}\thanks{Present address: Department of Electrical and Computer Engineering, Nagoya Institute of Technology, Nagoya 466-8555}, Shigeo \textsc{Ohara}, Masato \textsc{Hedo}$^{1}$, Yoshiya \textsc{Uwatoko}$^{1},$\\
Kazuya \textsc{Saito}$^{2}$, Michio \textsc{Sorai}$^{2}$ and Isao \textsc{Sakamoto}
}

\inst{%
Department of Electrical and Computer Engineering, Nagoya Institute of Technology, Nagoya 466-8555\\
$^1$The Institute for Solid State Physics, The University of Tokyo, Kashiwa 277-8581\\
$^2$Research Center for Molecular Thermodynamics, Graduate School of Science,\\Osaka University, Toyonaka, Osaka 560-0043\\
}

\recdate{\today}

\abst{%
We succeeded in growing a single crystal of Ce$_{2}$CoIn$_{8}$ by the flux method. The results of specific heat and electrical resistivity measurements indicate that Ce$_{2}$CoIn$_{8}$ is a heavy-fermion superconductor below 0.4 K with an electronic specific heat coefficient $\gamma$ as large as 500 mJ/K$^{2}$mol-Ce.
}

\kword{%
superconductivity, heavy fermion, crystal growth, specific heat, electrical resistivity, Ce$_{2}$CoIn$_{8}$
}

\begin{document}
\sloppy
\maketitle

Considerable research has recently been focused on a new family of Ce-based heavy-fermion compounds with the general formula Ce$_{\it n}${\it M}In$_{3{\it n}+2}$ ({\it M}=Co, Rh, Ir; {\it n}=1, 2), which crystallize in tetragonal Ho$_{\it n}$CoGa$_{3{\it n}+2}$-type structures. \cite {rf1,rf2,rf3} The crystal structure of Ce$_{\it n}${\it M}In$_{3{\it n}+2}$ can be seen as the {\it n} layers of CeIn$_{3}$ stacked sequentially along the {\it c}-axis with intervening one layer of {\it M}In$_{2}$. These compounds exhibit a variety of interesting phenomena, such as  heavy-fermion superconductivity, antiferromagnetism and pressure-induced superconductivity.

 For transition metals {\it M}=Rh and Ir,  a series of compounds {\it n}=1 and 2 has been synthesized in single crystal form. For {\it M}=Rh, both {\it n}=1 and 2 compounds are antiferromagnets with  N\'{e}el temperatures of {\it T}$_{\it N}$=3.8 and 2.8 K for CeRhIn$_5$ and  Ce$_{2}$RhIn$_{8}$, respectively.\cite {rf6,rf4} The electronic specific heat coefficient $\gamma$ is estimated to be about 50 mJ/K$^{2}$mol-Ce for CeRhIn$_5$ and 370 mJ/K$^{2}$mol-Ce for  Ce$_{2}$RhIn$_{8}$.\cite {rf12,rf13} In both compounds, superconductivity was observed under high pressures.  CeRhIn$_5$ transforms from antiferromagnetic state to superconducting state at 2 K under the pressure of 1.6 GPa.\cite {rf6}  For Ce$_{2}$RhIn$_{8}$, very recently, pressure-induced superconductivity was reported.\cite{rf14}

 For {\it M}=Ir, the {\it n}=1 member CeIrIn$_5$ is a heavy-fermion superconductor with a transition temperature of {\it T}$_c$=0.4 K at ambient pressure, while the {\it n}=2 member Ce$_{2}$IrIn$_{8}$ is a paramagnet down to 50 mK with no evidence of a phase transition. \cite {rf5,rf4} The $\gamma$ values are about 700 mJ/K$^{2}$mol-Ce for both compounds. \cite {rf4,rf16 }

For {\it M}=Co, the {\it n}=1 member CeCoIn$_{5}$ is a heavy-fermion superconductor with  a transition temperature of {\it T}$_c$=2.3 K. \cite {rf7}  The $\gamma$ value of CeCoIn$_{5}$  is about 300 mJ/K$^{2}$mol-Ce at {\it T}$_c$. When the superconductivity is suppressed by a magnetic field of 50 kOe, the $\gamma$ value increases with decreasing temperature and reaches a very large value of about 1 J/K$^{2}$mol-Ce. \cite {rf7, rf15} To our knowledge, however, there is no report on  the {\it n}=2 member Ce$_{2}$CoIn$_{8}$, except for the structural study.\cite {rf1,rf2}

The purpose of this paper is to clarify the thermal and electronic properties of Ce$_{2}$CoIn$_{8}$.  We attempted to synthesize Ce$_{2}$CoIn$_{8}$ by the flux method and found the conditions for single crystal growth. Measurements of specific heat and electrical resistivity were carried out on single crystals of Ce$_{2}$CoIn$_{8}$. In this paper, we report the observed superconductivity at {\it T}$_c$$\sim$0.4 K for Ce$_{2}$CoIn$_{8}$ at ambient pressure.

Single crystals of Ce$_{2}$CoIn$_{8}$ were grown from an In flux. An arc-melted Ce$_{2}$Co button and excess In were sealed in a quartz ampoule under high vacuum. The ampoule was heated up to 750$^\circ$C over 7 h and then raised quickly to 1000$^\circ$C and held for 5 minutes, followed by rapid cooling to 750$^\circ$C. The single crystals grew in the temperature region of 750-300$^\circ$C. After spinning off the remaining In, many mm-sized, cube(or tetragon)-with-octahedron-like single crystals were obtained. The residual In on grown crystal surfaces was further removed by acid etching. The crystals were characterized by a powder X-ray diffraction method using Cu{\it K}$\alpha$ radiation at room temperature. The specific heat was measured by the relaxation method in the temperature range of 2-20 K. The resistivity measurement was performed by a standard four-probe ac bridge method down to 0.1 K.

Figure 1 shows the X-ray diffraction pattern obtained for crushed crystals. The X-ray pattern is well indexed on the base of the tetragonal Ho$_{2}$CoGa$_{8}$-type structure (the space group P4/mmm), as indicated by the indices in Fig. 1. There is no trace of CeCoIn$_{5}$. A weak peak, marked by an arrow, is the (011) reflection of the residual In-flux. The lattice parameters are determined as {\it a} = 4.643 {\AA} and {\it c}  = 12.25 {\AA}, which agree with the reported values.\cite {rf1}

\begin{fullfigure}
\includegraphics[width=16cm,keepaspectratio,clip]{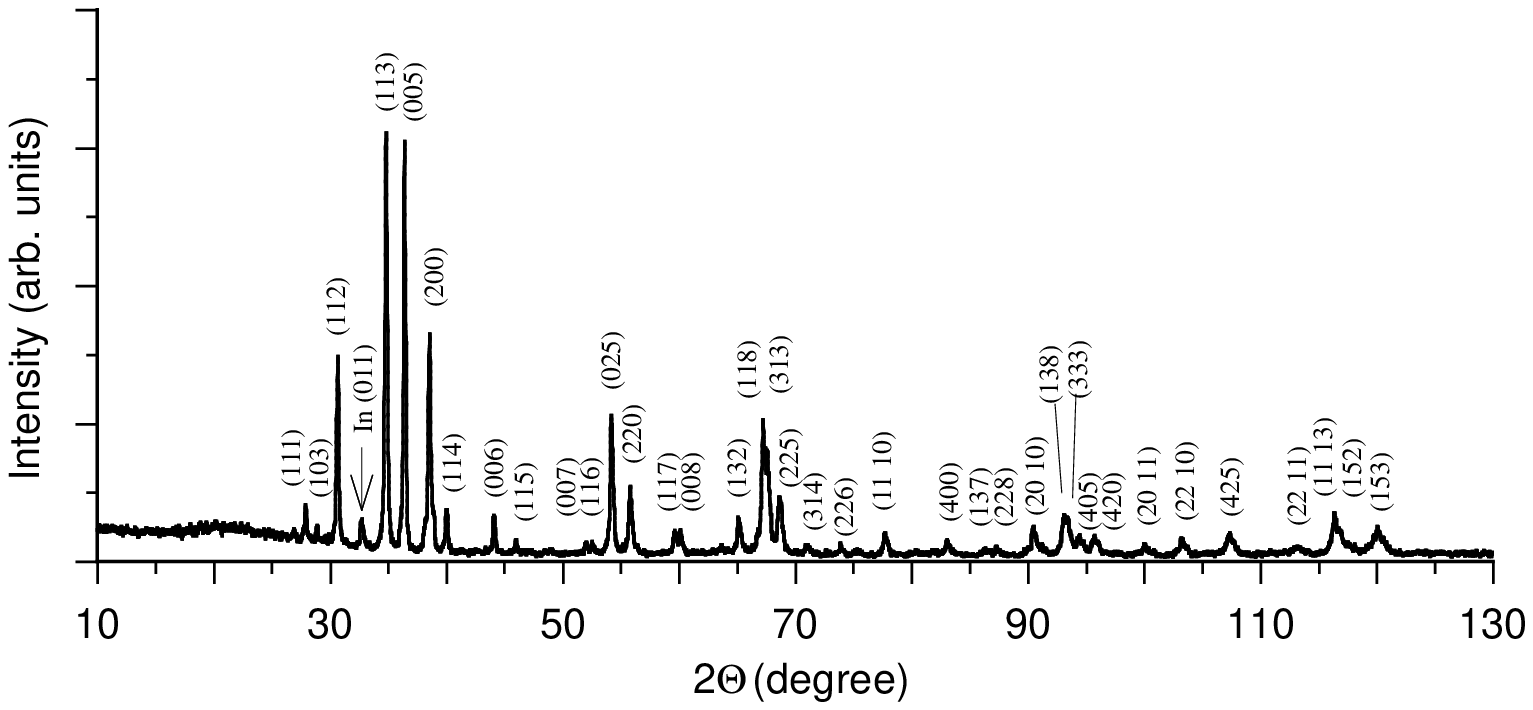}
\caption{X-ray diffraction pattern of Ce$_{2}$CoIn$_{8}$ with Miller indices. The arrow denotes the (011) reflection of the residual In-flux.}
\label{fig:1}
\end{fullfigure}
\begin{figure}
\begin{center}
\includegraphics[width=8cm,keepaspectratio,clip]{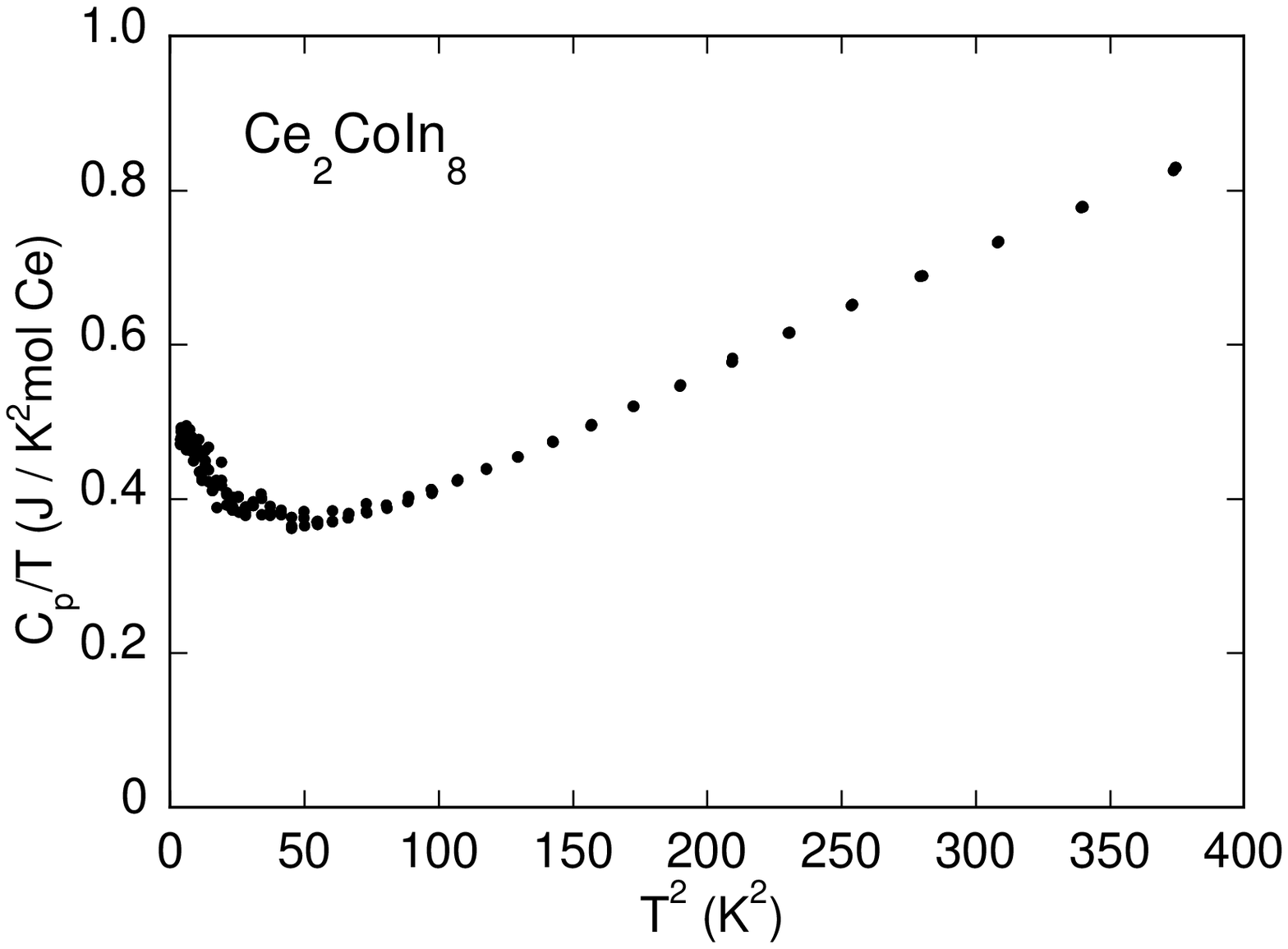}
\end{center}
\caption{Specific heat divided by temperature {\it C}$_p$/{\it T} versus {\it T}$^{2}$ for Ce$_{2}$CoIn$_{8}$. }
\label{fig:2}
\end{figure}
\begin{figure}
\begin{center}
\includegraphics[width=8cm,keepaspectratio,clip]{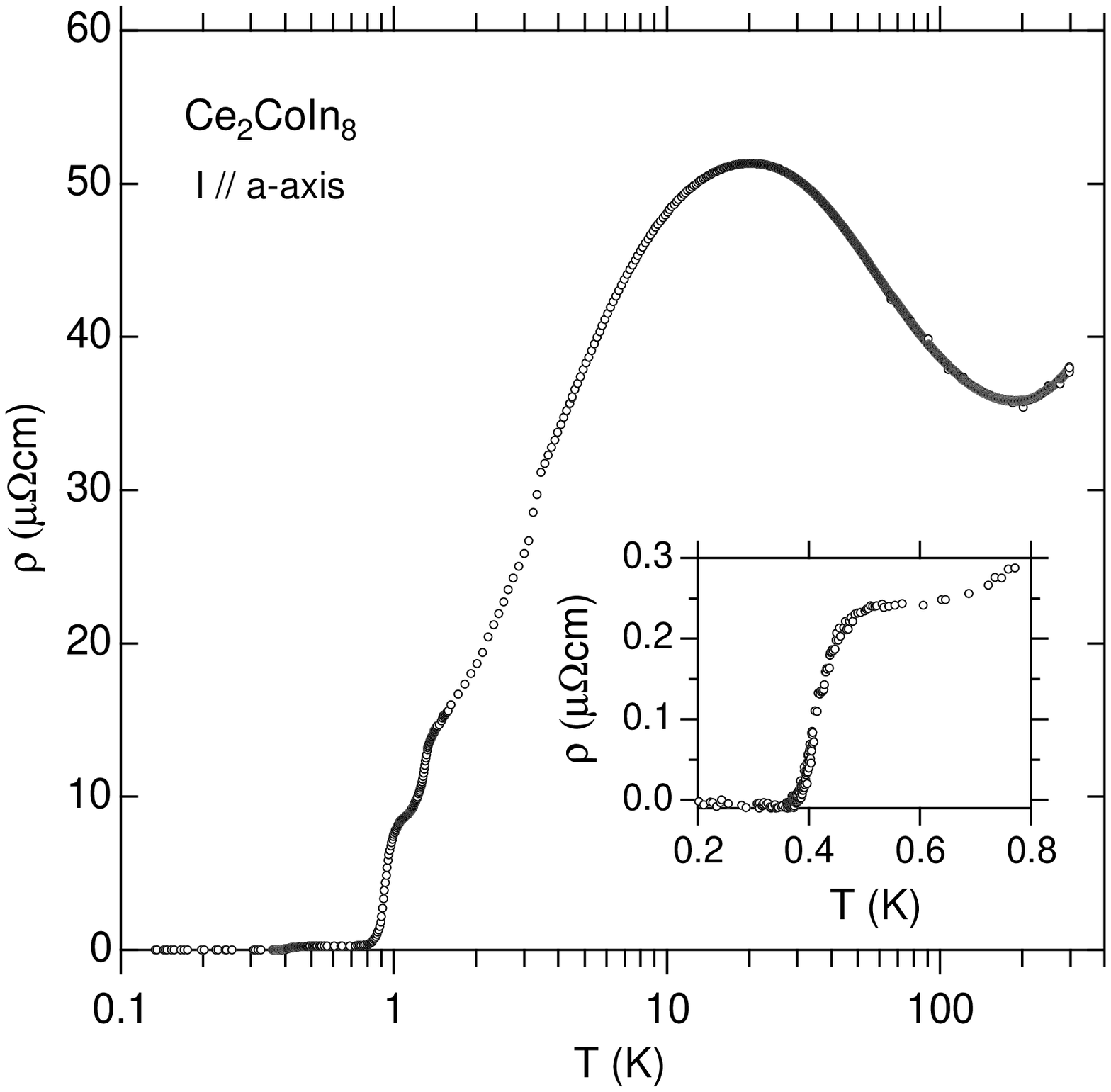}
\end{center}
\caption{Electrical resistivity plotted with log{\it T} in the temperature range from 0.1 to 300 K for Ce$_{2}$CoIn$_{8}$. The inset shows the resistivity in the lowest temperature region on a  linear scale.}
\label{fig:3}
\end{figure}
{{
Specific heat divided by temperature {\it C}$_{\it p}$/{\it T} {\it vs} {\it T}$^{2}$ for Ce$_{2}$CoIn$_{8}$ is illustrated in Fig. 2.  With decreasing temperature, {\it C}$_{\it p}$/{\it T} decreases linearly for {\it T}$^{2}$ and shows a minimum around 7 K. At lower temperatures, {\it C}$_{\it p}$/{\it T} increases with decreasing temperature. At 2 K,  {\it C}$_{\it p}$/{\it T} reaches 500 mJ/K$^{2}$mol-Ce.  This indicates that the $\gamma$ value of Ce$_{2}$CoIn$_{8}$ is more than 500 mJ/K$^{2}$mol-Ce, showing a heavy-fermion characteristic.

Figure 3 shows the temperature dependence of the electrical resistivity $\rho$ as a function of log {\it T} with the current along the {\it a}-axis. The inset shows the resistivity below 0.8 K on a linear scale. With decreasing temperature, the resistivity shows a -log {\it T} dependence below 80 K down to 30 K, reaches a maximum, and decreases at lower temperatures. A kink at 3.4 K is attributed to the superconducting transition of free In that might be included in the crystal. 
As the temperature decreases further, the $\rho$-{\it T} curve shows a few abrupt drops near 1.4 K, 1.0 K and 0.4 K, as shown in Fig. 3. Below 0.4 K, the resistivity becomes zero, as shown in the inset in Fig. 3. This suggests  that Ce$_{2}$CoIn$_{8}$ is a heavy-fermion superconductor with a transition temperature of {\it T}$_c$$\sim$0.4 K.

To determine the upper critical field {\it H}$_ {c2}$ for Ce$_{2}$CoIn$_{8}$, we measured the resistivity on a single crystal Ce$_{2}$CoIn$_{8}$ with a magnetic field applied along the {\it a}-axis. The zero resistivity state is broken with the field  {\it H}=1 T at 0.1 K, which indicates that {\it H}$_{c2}$$\sim$1 T for Ce$_{2}$CoIn$_{8}$.  The values of {\it T}$_c$$\sim$0.4 K and {\it H}$_{c2}$$\sim$1 T  for Ce$_{2}$CoIn$_{8}$ are much lower than those for  CeCoIn$_{5}$, but are very similar to those for CeIrIn$_{5}$.\cite {rf12,rf9,rf10} The complete {\it H}-{\it T} phase diagram will be reported in the near future.

The causes of the resistivity drops near {\it T}=1.4 and 1.0 K for Ce$_{2}$CoIn$_{8}$ are yet to be clarified. However, a similar resistivity anomaly is also observed in CeIrIn$_{5}$, where the anomaly is viewed as an appearance of superconductivity but is distinguished from bulk superconductivity.\cite {rf5,rf12,rf17}  To clarify the origin of these resistivity drops and the bulk properties of the superconducting state for Ce$_{2}$CoIn$_{8}$, more detailed studies of specific heat and magnetic susceptibility at lower temperatures are being carried out.

Here, we have presented the specific heat and the resistivity measurements performed on the single crystals of Ce$_{2}$CoIn$_{8}$. It is found that Ce$_{2}$CoIn$_{8}$ is a heavy-fermion superconductor with {\it T}$_c$$\sim$0.4 K.

We thank S. Okada, Y. Shomi and Y. X. Jin for their experimental assistance.  One of the authors (S. Ohara) was financially supported by the Nitto Foundation.

\end{document}